\documentclass[floats,floatfix,showpacs,superscriptaddress,nofootinbib,twocolumn]{revtex4}

\usepackage{amsmath,amssymb,epsfig,url,balance}
\usepackage[tight]{subfigure}

\newcommand{\BEQA}{\begin{eqnarray}}
\newcommand{\EEQA}{\end{eqnarray}}

\newtheorem{validation}{Validation Against Observed Topology}

\begin{document}

\title{Evolution of the Internet AS-Level Ecosystem}
\author{Srinivas Shakkottai}
\affiliation{
Department of ECE, Texas A\&M University,
College Station, TX 77843-3128, USA}
\author{Marina Fomenkov}
\author{Ryan Koga}
\author{Dmitri Krioukov\thanks{dima@caida.org}}
\author{kc claffy}
\affiliation{
Cooperative Association for Internet Data Analysis, University of California, San Diego,
La Jolla, CA 92093-0505, USA}
\begin{abstract}
We present an analytically tractable model of Internet evolution at
the level of Autonomous Systems (ASs). We call our model the
multiclass preferential attachment (MPA) model. As its name
suggests, it is based on preferential attachment. All of its
parameters are measurable from available Internet topology data.
Given the estimated values of these parameters, our analytic results
predict a definitive set of statistics characterizing the AS
topology structure. These statistics are not part of the model
formulation. The MPA model thus closes the
``measure-model-validate-predict'' loop, and provides further
evidence that preferential attachment is a driving force behind
Internet evolution.
\end{abstract}
\maketitle

\section{Introduction}

The Internet is a paradigmatic example of a complex network.  Many
researchers have used publicly available data on Internet topology
and its observed evolution to test a variety of physical models of
complex network structure and dynamics.
The large-scale Internet topology represent the structure
of connections between companies owing parts of the Internet infrastructure,
each company roughly corresponding to an {\em Autonomous System~(AS)}~\cite{PaSaVe04-book}.
An AS might be a transit Internet service provider
(ISP) or customer, a content provider or sink, or any combination
of these. Some ASs span multiple continents and are highly
interconnected, while others are present at a single geographical
location and have only a few links. In 1999 Faloutsos {\em et
al.}~\cite{FaFaFa99} observed that despite all this diversity,
the distribution of AS degrees obeys a simple power law. This
observation remains valid today, after ten years of Internet
evolution~\cite{1452543}.

Many researchers have attempted to model the Internet as an evolving
system~\cite{AlBa00,YoJeBa02,GoKaKi02,944780,ZhoMo04,Zhou06,SeBoDi05,SeBoDi06,DBLP:conf/infocom/ChangJW06,DBLP:conf/infocom/WangL06,SoBo07,CoJa07,BaGoWo07,HoKa08},
and have studied its properties~\cite{PaVaVe01,RaBa03,PaNe03,BiCaCa05}.
However, questions regarding the main drivers behind Internet
topology evolution remain~\cite{wit-ccr}. In this paper our main
objective is to create an evolutionary model of the AS-level
Internet topology that simultaneously:
\begin{enumerate}
\item is realistic,
\item is parsimonious,
\item has {\em all\/} of its parameters measurable,
\item is analytically tractable, and
\item ``closes the loop.''
\end{enumerate}
Parsimony implies that the model should be as simple as possible,
and, related to that, the number of its parameters should be as
small as possible. The fifth requirement means that if we substitute
measured values of these parameters into analytic expressions of the
model, then these expressions will yield results matching empirical
observations of the Internet. However, most critical is the third
requirement~\cite{wit-ccr}: as soon as a model has even a few
unmeasurable parameters, one can freely tune them to match
observations. Such parameter tweaking to fit the data may create an
illusion that the model ``closes the loop,'' but in the end it
inevitably diminishes the value of the model because there is no
rigorous way to tell why one model of this sort is better than
another since they all match observations. To the best of our
knowledge, the multiclass preferential attachment (MPA) model that
we propose and analyze in this paper is the first model satisfying
all five requirements listed above.

A salient characteristic of our model is that we distinguish between
two kinds of ASs: ISPs and non-ISPs. The main difference between
these two types of ASs is that while both ISPs and non-ISPs can
connect to ISPs, no new AS will connect to an existing non-ISP since
the latter does not provide transit Internet connectivity. In
Section \ref{sec_2cls_prf_attach} we analyze the effect of this
distinction on the degree distribution.
In Section \ref{sec_2nd_order} we account for other processes. ISPs
can form peering links to exchange traffic bilaterally. They can
also go bankrupt and be acquired by others. Finally, they can
multihome, i.e., connect to multiple providers. Prior work has often
focused on these processes as the driving forces behind Internet
topology evolution. We show that in reality they have relatively
little effect on the degree distribution. Using the best available
Internet topology data, we measure the parameters reflecting all the
process above, and analytically study how they affect the degree
distribution.

However, the degree distribution alone does not fully capture the
properties of the Internet AS graph~\cite{MaKrFaVa06-phys}. The
$dK$-series formalism introduced in~\cite{MaKrFaVa06-phys} defines a
systematic basis of higher-order degree distributions/correlations.
The first-order ($1K$) degree distribution reduces to a traditional
degree distribution. The second-order ($2K$) distribution is the
joint degree distribution, i.e., the correlation of degrees of
connected nodes. The distributions can be further extended to
account for higher-order correlations~\cite{MaKrFaVa06-phys}, or for
different types of nodes and links, called
annotations~\cite{DiKr09}. In the economic AS Internet, there are
two types of links: links connecting customer ASs to their providers
(customer-to-provider links), and links connecting ISPs to their peers (peer-to-peer links).
Reproducing the $2K$-annotated distribution of AS topologies
suffices to accurately capture virtually all important topology
metrics~\cite{MaKrFaVa06-phys,DiKr09}. An important feature of the
MPA model is that, by construction, it naturally annotates the links
between ASs by their business relationships, and that we can
analytically calculate, in Section~\ref{sec:ADs}, the distributions
for the number of peers, customers, and providers that nodes have.
In Section \ref{sec_sims} we perform a $2K$-annotated test. We
generate synthetic graphs using the MPA model, and find that these
graphs exhibit a startling similarity to the observed AS topology
according to almost all the $2K$-annotated statistics. This
validation, in conjunction with observations
in~\cite{MaKrFaVa06-phys,DiKr09}, ensures that other important
topology metrics also match well.

\section{Two-Class Preferential Attachment Model}\label{sec_2cls_prf_attach}

In the original preferential attachment (PA) model~\cite{BarAlb99},
there is only one type of nodes.  Suppose that nodes arrive in the
system at the rate of one node per unit time, and let them be
numbered $s=1,2,3,...$ as they arrive. Then the number of nodes in
the system at time $t$ is equal to $t$.  A node entering the system
brings a link with it.  One end of the link is already connected to
the entering node, while the other end is loose.  We call such an
un-associated end a \emph{loose connection}.  According to the PA
model, nodes attach to existing ones with a probability proportional
to their degrees. Thus, the probability that a node of degree $k$ is
selected as a target for the incoming loose connection, is its
degree divided by the total number of existing connections in the
system, which is $\frac{k}{2t}$. The original PA model yields a
power-law degree distribution $P(k)\sim k^{-\gamma}$ with
$\gamma=3$, but the linear preference function can be modified by an
additive term such that the model produces power laws with any
$\gamma>2$~\cite{BarAlb99,DoMeSa00,KraReLe00}.

\begin{figure}
\begin{center}
\epsfig{file=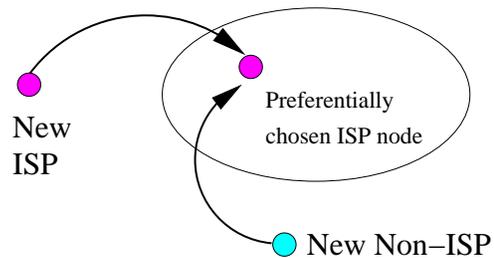,width=2.5in}
\caption{Two-class preferential attachment model.  There are ISP nodes and non-ISP nodes.}
\label{fig_two_cls_preferential}
\end{center}
\vskip -6pt
\end{figure}

In the Internet, there are two fundamentally different types of
ASs---ISPs and non-ISPs---that differ in whether they provide
traffic carriage between the ASs they connect or not. No new AS
would connect to an existing non-ISP since it cannot provide
Internet connectivity.  No other work has attempted to model this
observation, which is fundamental to understanding the evolution of
Internet AS-level topology.  Thus, our first modification of the PA
model is to consider these two classes of nodes (Figure
\ref{fig_two_cls_preferential}). New ISPs appear at a rate $1$ and
connect to other ISP-nodes with a linear preference. New non-ISPs
appear at some rate $\rho$ per unit time, implying
that the ratio $r$ of non-ISP nodes to the total number of nodes is
\BEQA\nonumber
r=\frac{\rho}{1+\rho}.
\EEQA
Non-ISPs attach to
existing ISP-nodes with a linear preference. However, \emph{no
further attachments to non-ISPs can occur}. Thus, with respect to
degree distribution, only the ISP-nodes contribute to the tail of
the power law as the non-ISP nodes will all have degree $1$. Given
that a link between an ISP node and a non-ISP node has only one end
that contributes to the degree of ISP nodes, and since we look only
at ISP nodes to find the degree distribution, the links that have an
ISP node on one end and a non-ISP node on the other should be
counted only once, i.e., they are counted as contributing $1$
connection.

Since each ISP node contributes $2$ connections to the network and a non-ISP node contributes only $1$, the total number of existing connections in the network at
time $t$ is $(2 + \rho )t$, which implies that the probability of any loose connection connecting to an ISP-node of
degree $k$ is
\BEQA\label{eqn_prob}
\frac{k}{(2+\rho)t }.
\EEQA
We use the notation $p(k,s,t)$ to denote the probability that an ISP node $s$ has degree $k$ at time $t$.  Then the average degree of an ISP node $s$ at time $t$ is
\BEQA
\bar{k}(s,t) = \sum_k k p(k,s,t) \EEQA
Both entering ISPs and non-ISPs have one loose connection each, so the number of loose connections entering the system at time $t$ is $1+\rho$.  Then, from
(\ref{eqn_prob}), the continuous-time model of the system is
\BEQA\label{eqn_evolv3}
\frac{\partial\bar{k}(s,t)}{\partial t} = \frac{1+\rho}{(2+\rho)t} \bar{k}(s,t),
\EEQA
with boundary condition $\bar{k}(t,t)=1$ for $t>1$.  Solving this equation yields
\BEQA\label{eq:deg-1}
\bar{k}(s,t)=\left(\frac{s}{t}\right)^{-\frac{1+\rho}{2+\rho}}.
\EEQA
This model represents a deterministic system in which if ISP node $s$ has degree $\bar{k}$, then ISP nodes that arrived before $s$ (in the interval $[0,s)$) have
degree at least $\bar{k}$.  Thus, $s$ also represents the number of ISP nodes that have degree at least $\bar{k}$. It follows from (\ref{eq:deg-1}) that the number of
ISP nodes that have degree $\bar{k}$ or higher is $$\frac{t}{\bar{k}^{\frac{2+\rho}{1+\rho}}}.$$  Note that the number of ISP nodes that arrive in $[0,t]$ is just
$t$.  Hence, the \emph{fraction} of ISP nodes that have degree $\bar{k}$ or higher is
$$\frac{1}{\bar{k}^{\frac{2+\rho}{1+\rho}}}.$$
Since this fraction is essentially a complimentary cumulative distribution function (CCDF), we differentiate it and multiply by $-1$ to obtain the density function
\BEQA
f(\bar{k})=\frac{2+\rho}{1+\rho}{\bar{k}^{-\left(2+\frac{1}{1+\rho}\right)}},
\EEQA
which corresponds to the probability distribution
\BEQA\label{eq_two_cls_expnt}
P(k)\sim k^{-\left(2+\frac{1}{1+\rho}\right)}=k^{-\left(3-r\right)}.
\EEQA

\begin{validation}
Dimitropoulos \emph{et~al.}~\cite{DiKrRi06} applied machine learning
tools to the best available data from the Internet registry (WHOIS)
and routing (Border Gateway Protocol (BGP)) systems to classify ASs
into several different classes, such as Tier-1 ISPs, Tier-2 ISPs,
Internet exchange points, universities, customer ASs, and so on.
They validated the resulting taxonomy by direct examination of a
large number of ASs. We use their results and divide ASs into two
classes based on whether they are ISPs or not. According
to~\cite{DiKrRi06} (the dataset is available at
\url{http://www.caida.org/data/active/as_taxonomy/}),
the number of ISPs is about $30\%$ of ASs, while
non-ISPs make up the other $70\%$, i.e., $r=0.7$ and $\rho=7/3$.
The measured value of $\rho$ yields $P(k)\sim k^{-2.3}$, with exponent close to the
observed value between $-2.1$ and
$-2.2$~\cite{FaFaFa99,ChaGoJaSheWi04,MaKrFo06}.
\end{validation}

The key point of this section is that the observed value of the
power-law exponent close to $2$ finds a natural and simple explanation: it is due
to preferential attachment and to a directly measured high
proportion of non-ISP nodes to which newly appearing nodes cannot
connect.

\section{\label{sec_2nd_order}Multiclass Preferential Attachment: Peering, Bankruptcy, Multihoming, and Geography}

In this section we add further refinements to our model and show
that, contrary to common beliefs, none of these refinements have a
significant impact on the degree distribution shape.

Relationships between ASs change over time, as ASs pursue
cost-saving measures. If the magnitude of traffic flow between two
ISPs is similar in both directions, then reciprocal peering with
each other allows each ISP to reduce its transit costs.  Under the
assumption that all customer ASs generate similar volumes of
traffic, high degree ASs would exchange high traffic volume and
rationally seek to establish reciprocal peering with other high
degree ASs.  We denote the rate at which peering links appear by
$c$.  The probability that a new peering link becomes attached to a
pair of ISP-nodes of degree $k_1$ and $k_2$ is proportional to
$k_1k_2$.

When ISPs go bankrupt, their infrastructure is usually acquired by
another ISP, which then either merges the ASs or forms a ``sibling''
relationship in which their routing domains appear independent but
are controlled by one umbrella organization. Thus, in terms of the
topology, bankruptcy means that a connection shifts from one ISP to
another. Since high degree ISPs tend to be wealthier, they are more
likely to be involved in such takeovers. We denote the rate of
bankruptcy by $\mu$ per unit time.

\begin{figure}
\begin{center}
\epsfig{file=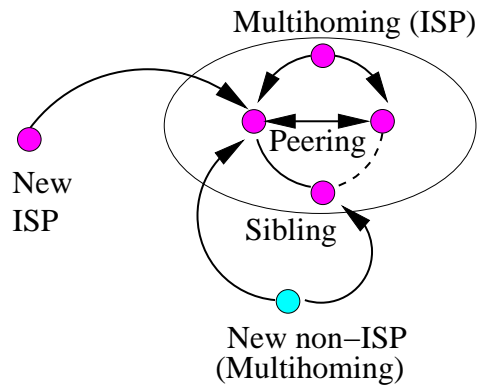,width=2.5in}
\caption{The multiclass preferential attachment model.}
\label{fig_complete2}
\end{center}
\vskip -6pt
\end{figure}

A growing AS may decide to multihome, i.e., to connect to at least
two Internet providers. One would expect that higher degree ISPs
with a need for reliability would multihome to other higher degree
ISPs.  We model this phenomenon by assuming that multihoming links
appear in the system at rate $\nu$ per unit time.  The probability
that a new link becomes attached to a pair of ISP-nodes of degree
$k_1$ and $k_2$ is proportional to $k_1k_2$. The links are directed
from the customer to the provider, and we assume that the higher
degree ISP is the provider.  We also assume that non-ISPs multihome
to an average of $m$ providers each. The model is illustrated in
Figure~\ref{fig_complete2}.

We analyze this complete MPA model, using techniques similar to that of Section \ref{sec_2cls_prf_attach}.  The relation we obtain for $\bar{k}(s,t)$ is
\BEQA\label{eqn_final_soln}
\bar{k}(s,t) =
\left(\frac{s}{t}\right)^{-\alpha} + \frac{\mu}{\alpha},
\EEQA
where
\BEQA\label{eqn_final_alpha}
\alpha=\frac{1+2\nu + m\rho + 2c +m\mu}{2+2\nu +m\rho + 2c}.
\EEQA
Proceeding in a manner identical to Section \ref{sec_2cls_prf_attach} yields the relation
\BEQA
P(k)\sim k^{-\gamma},
\EEQA
where
\BEQA\label{eqn_gamma_alpha}
\gamma = \frac{1}{\alpha}+1 =  2 + \frac{1-\mu}{1 + 2\nu +m \rho +2c + \mu}.
\EEQA

\begin{validation}
We used the annotated Route Views data~\cite{routeviews} from~\cite{DiKrFo06}
(the dataset is available at \url{http://www.caida.org/data/active/as_taxonomy/})
in order to obtain the empirical distribution of
number of ISPs to which ASs multihome. We find that the average
number of providers that ISPs connect to is $2$,
meaning that $\nu=1$. Indeed, since
ISPs arrive at rate $1$ and choose one provider initially,
multihoming links entering at rate $\nu = 1$ yield the
average number of providers per ISP of $2$.
The average number of providers that non-ISPs multihome to is $1.86$,
i.e., $m=1.86$. Dimitropoulos \emph{et. al} \cite{DiKrFo06} also
showed that roughly $90\%$ of links are of customer-provider type,
i.e., these links pertain to transit relationships, with payments
always going to the provider ISP. They find (a lower bound of)
$10\%$ of links are peering, i.e., these links correspond to
bilateral traffic exchange without payment. In the model, customer
links appear in the system at a rate of $1 + m\rho$. We thus
calculate $c =(1 + \nu + m\rho)/9 = 0.704$ peering links per unit
time.  The authors of~\cite{DiKrFo06} also estimate that the
fraction of sibling links is too small to measure accurately and we
take $\mu=0$. Substituting these values into the exponent expression
(\ref{eqn_gamma_alpha}) results in $\gamma=2.114$ that matches the
observed values lying between $2.1$ and
$2.2$~\cite{FaFaFa99,ChaGoJaSheWi04,MaKrFo06}.
\end{validation}

The key point of this section is that the large ratio of non-ISPs to ISPs
$\rho$ is the dominating term in determining the value of $\gamma$,
bringing it down from $3$ to $2.3$,
while all the other parameters are less significant,
decreasing $\gamma$ further from $2.3$ to $2.1$, its observed value.
However, the extensions considered in this section do strongly affect other network
properties, such as clustering, as we will see in Section~\ref{sec_sims}.

Our last comment in this section concerns geography. We could divide
the world into different geographical regions, each growing at a
different rate. Due to the self-similar nature of scale-free
topologies~\cite{SeKrBo08}, the resulting graph would still bear
identical properties to the MPA model as long as the parameters
$\rho$, $c$, and $\mu$ are the same in all regions. Evidence
supporting this hypothesis is available in~\cite{ZhoZha07,MaKrFo06},
where Chinese or European parts of the Internet are shown to have
properties virtually identical, after proper rescaling, to the
global AS topology.

\section{Peer, Customer, and Provider Distributions}\label{sec:ADs}

In this section we calculate the distributions for the number of peers, customers, and providers that ISPs have in the MPA model.

The first two distributions are power laws with the exponents equal to $\gamma$, the exponent of the overall degree distribution. To see this, we first focus on the
peer distribution. We denote the average number of peers that an ISP arrived at time $s$ has at time $t$ by $\bar{\zeta}(s,t)$.  The dynamics of peer link formation
is given by
\BEQA
\frac{\partial\bar{\zeta}(s,t)}{\partial t} = \frac{2c}{(2+ 2\nu+m\rho+2c)t} \bar{k}(s,t),
\EEQA
since $2c$ is the rate of arrival of loose peer connections, which attach to target nodes with probability proportional to the target node degree.
If we define
\BEQA
\beta = \frac{2c}{(2+ 2\nu+m\rho+2c)}, \nonumber
\EEQA
so that,
\BEQA
\frac{\partial\bar{\zeta}(s,t)}{\partial t} = \frac{\beta}{t} \bar{k}(s,t),
\EEQA
then after the substitution of $\bar{k}(s,t)$ from (\ref{eqn_final_soln}) with $\mu =0$, we have
\BEQA
\frac{\partial\bar{\zeta}(s,t)}{\partial t} = \frac{\beta}{t}\left(\frac{s}{t}\right)^{-\alpha}.
\EEQA
We then solve the above for $\bar{\zeta}(s,t)$:
\BEQA
\bar{\zeta}(s,t)= \frac{\beta}{\alpha}\left(\frac{s}{t}\right)^{-\alpha}.
\EEQA
Thus, the number $s$ of ISP nodes that have $\bar{\zeta}$ or more peers is
\BEQA
s= \frac{t}{\left(\frac{\alpha}{\beta} \bar{\zeta}(s,t)\right)^{\frac{1}{\alpha}}}.
\EEQA
Dividing the right side by $t$ (the total number of ISP nodes in the system at time $t$) gives the cumulative distribution of the average number of peers of ISPs.
Differentiating and multiplying by $-1$ yields the distribution
\BEQA
f(\bar{\zeta})= \frac{\bar{\zeta}^{-\left(\frac{1}{\alpha} +1\right)}}{\alpha\left(\frac{\alpha}{\beta}\right)^{\frac{1}{\alpha}}},
\EEQA
which corresponds to $P(\zeta)\sim \zeta^{-\gamma}$ with the same power law exponent $\gamma=\frac{1}{\alpha} +1$ as in (\ref{eqn_gamma_alpha}).  We can show in an
identical fashion that the distribution for the number of customers that ISPs have follows the same power law distribution. We will check the validity of these
results in Section~\ref{sec_sims}.

We next show that the distribution for the number of providers that ISPs have is a random variable $1+X$, where $X$ approximately follows an exponential distribution
with parameter $\nu$.  According to the model, multihoming links between ISPs are directed from the lower to higher degree ISP.  Although the probability that the
multihoming link connects to a particular ISP is proportional to the ISP's degree, the probability that the chosen ISP is the customer end of the multihoming link is
inversely proportional to its degree.  We can thus expect the probability that any particular ISP obtains an extra provider at any time to be approximately the same.
Since upon its arrival each ISP always chooses a provider, the minimum number of providers of ISPs is $1$. If we now denote by $1+\bar{p}(s,t)$ the average number of
providers that ISPs, appeared at time $s$, have at time $t$, then $\bar{p}(s,t)$ is a solution of the following equation
\BEQA
\frac{\partial \bar{p}(s,t)}{\partial t} = \frac{\nu}{t},
\EEQA
because the rate of arrival of multihoming links is $\nu$.
Solving the above expression with the initial condition $\bar{p}(s,s)=0$,
we obtain
\BEQA
s=t e^{\frac{-\bar{p}}{\nu}}.
\EEQA
Since the number of ISPs at time $t$ is just $t$, this expression implies that the distribution of the number of ISPs that have $\bar{p}$ or more multihoming links is
exponential $e^{\frac{-\bar{p}}{\nu}}$.  Since all ISPs (except the first) have at least one provider, the distribution for the number of providers that ISPs have
should approximately be a random variable $1+X$, where $X$ is exponentially distributed with parameter $\nu$. Given that the argument above is rather heuristic, we
can hardly expect this distribution in the real Internet to be exactly exponential. However, the most important consequence of this argument is that this distribution is quite unlikely to be heavy-tailed.

\begin{validation}
Using the same data from~\cite{DiKrFo06}, the complementary
cumulative distribution function (CCDF) for the number of providers
that ISPs have (after subtracting the one initial provider) versus
the ISP degree is shown in Figure~\ref{fig_ISP_prov_ccdf} plotted in
the semi-log scale. The exponential curve fit to the initial part of
the graph has a slope of $-0.7$,  i.e., the average number of
providers is $1+1/0.7 = 2.4$, which is close to our empirically
measured mean value of $2$ in Validation~2.
\begin{figure}
\begin{center}
\epsfig{file=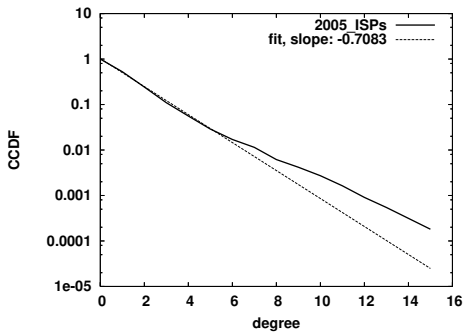,width=2.5in}
\caption{Provider distribution.} \label{fig_ISP_prov_ccdf}
\end{center}
\end{figure}
\end{validation}
The purpose of this validation is to show not that the distribution
is exactly of the form $1+X$ where $X\sim \exp(1/\nu)$, but that it
is definitely not a power law.

\section{Model Validation by Simulation}\label{sec_sims}

\begin{figure*}
\centering
\subfigure[Degree distribution]{
\epsfig{file=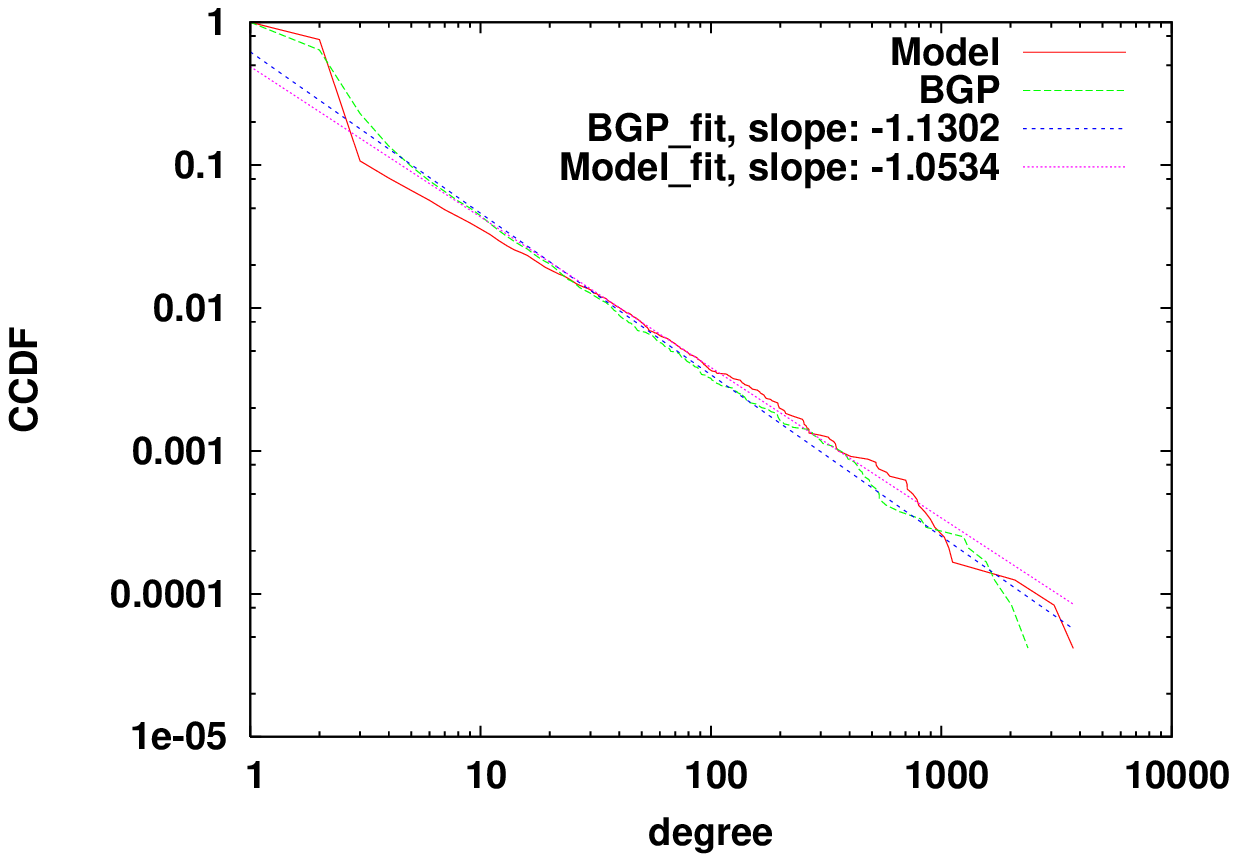,width=2.6in}
 \label{fig_all_ccdf}
}
\subfigure[Customer distribution]{
\epsfig{file=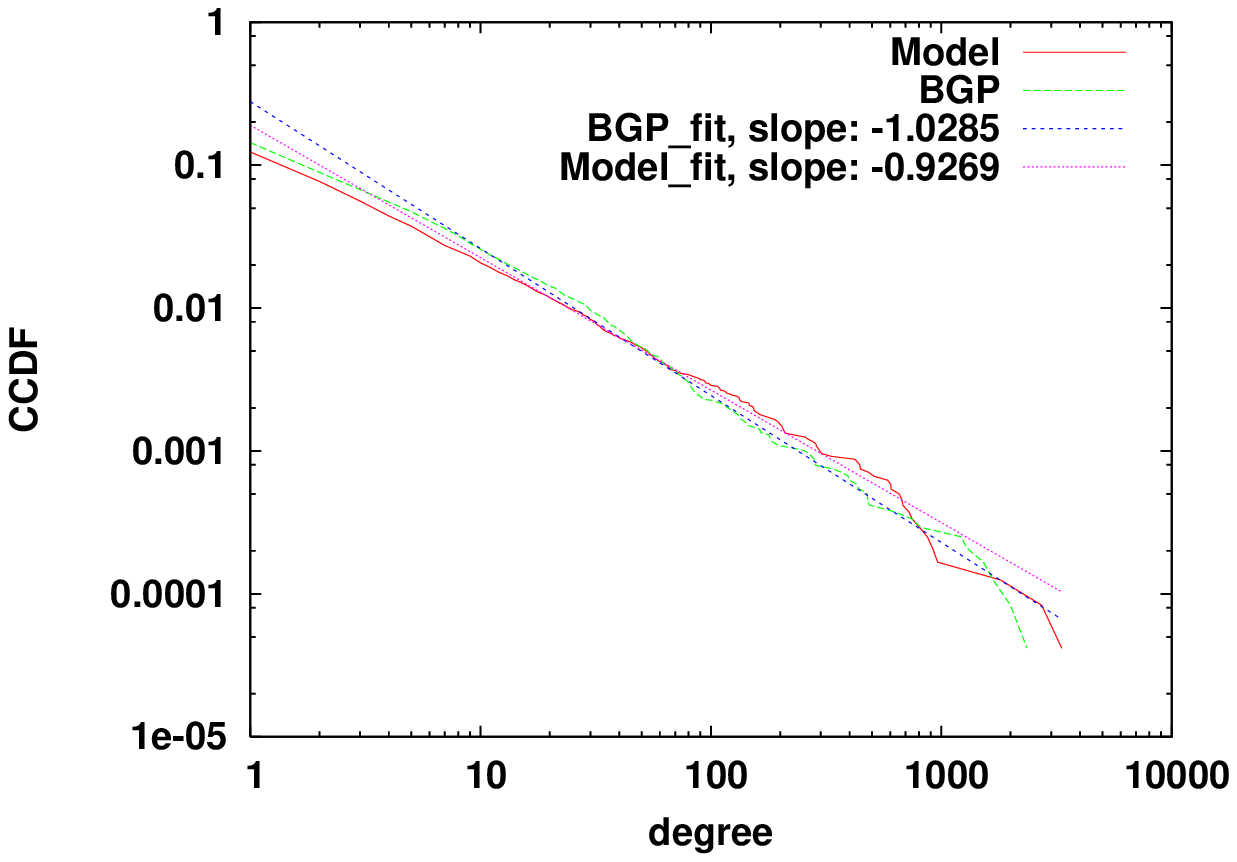,width=2.6in}
\label{fig_cust_ccdf}
}
\subfigure[Peer distribution]{
\epsfig{file=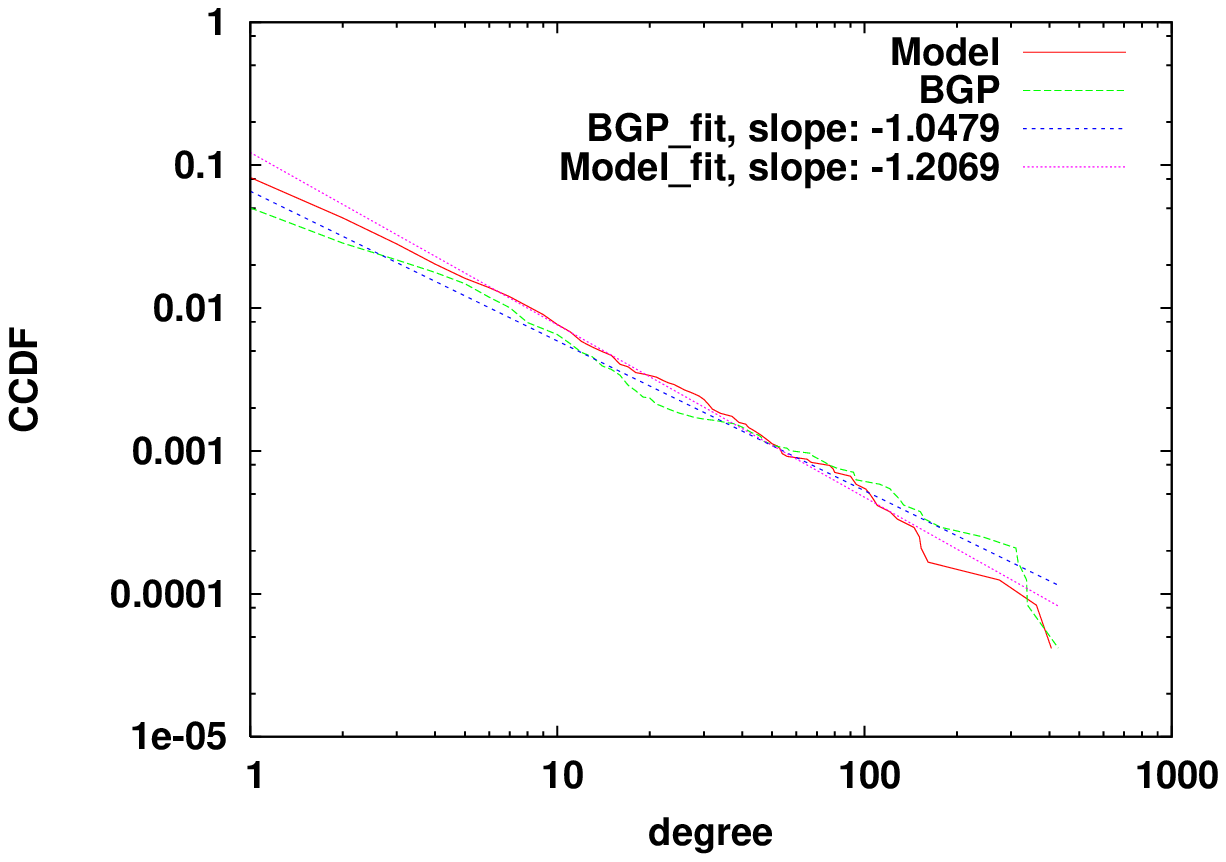,width=2.6in}
\label{fig_peer_ccdf}
}
\subfigure[Provider distribution]{
\epsfig{file=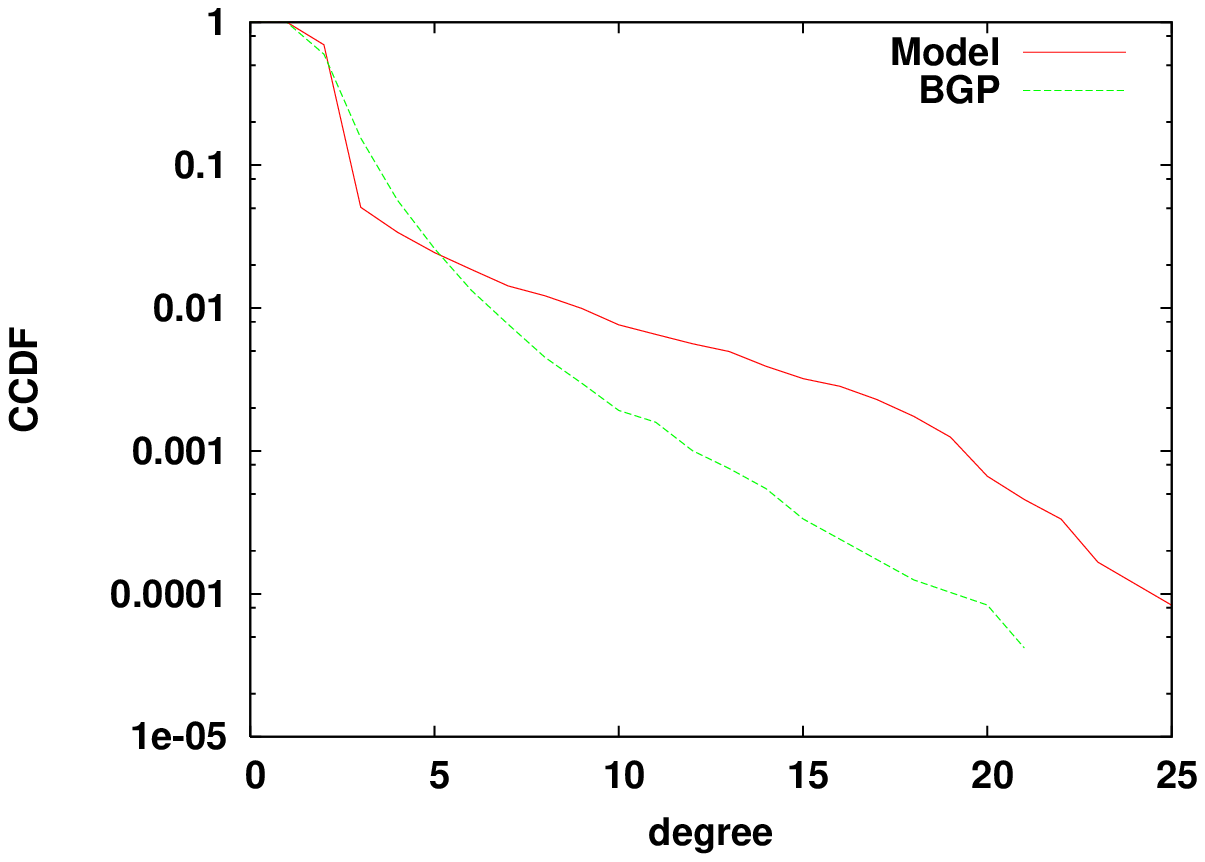,width=2.6in}
\label{fig_prov_ccdf}
}
\subfigure[Provider vs.\ cust.\ degree]{
\epsfig{file=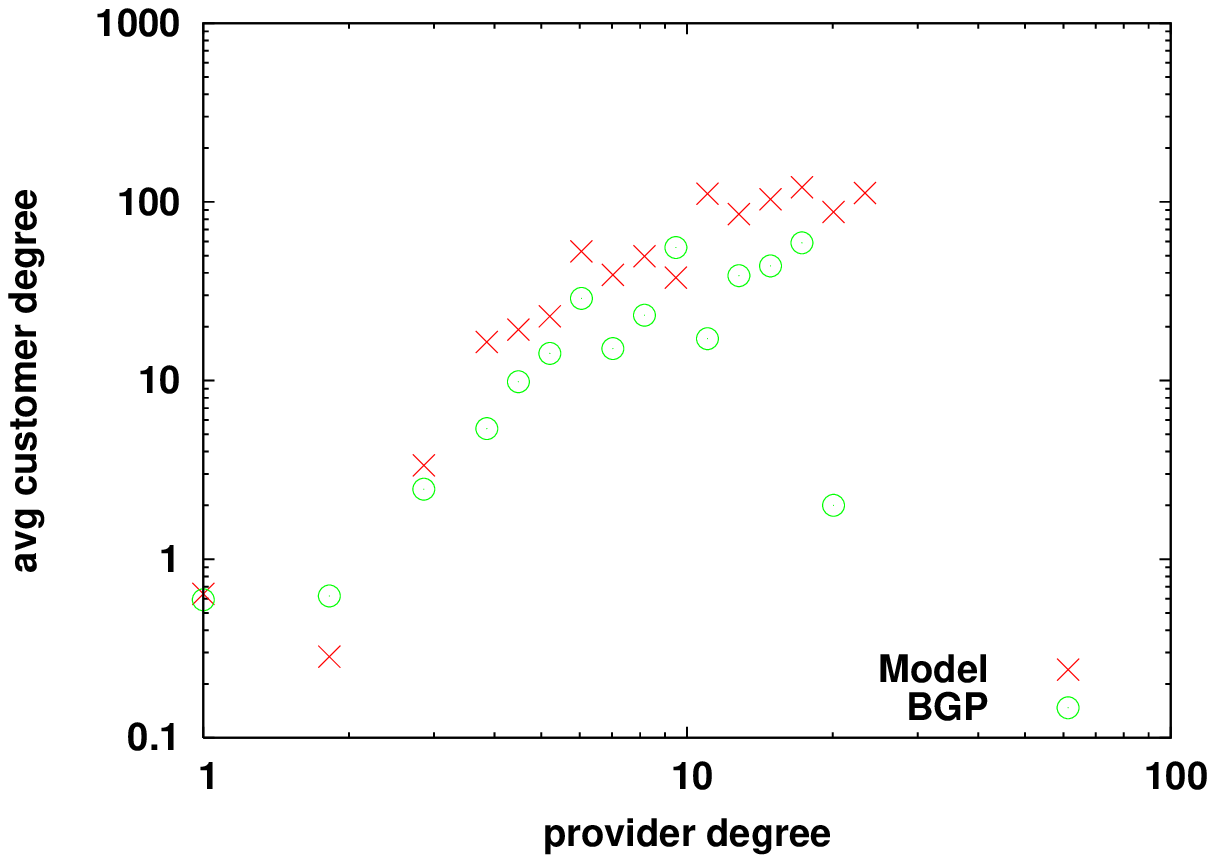,width=2.6in}
\label{fig_1_node_cust_vs_prov}
}
\subfigure[Provider vs.\ peer degree]{
\epsfig{file=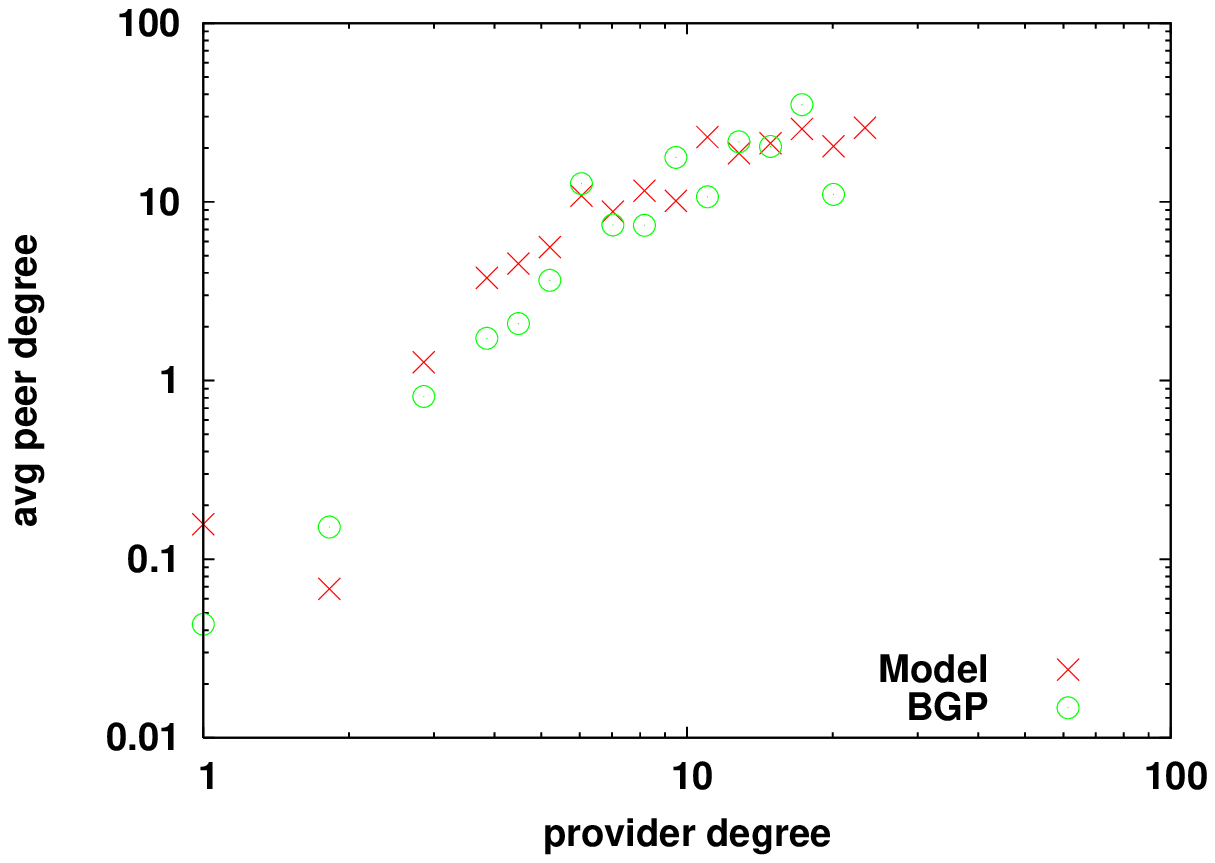,width=2.6in}
\label{fig_1_node_peer_vs_prov}
}
\subfigure[Node total degree vs.\ the average degree of its customers and providers]{
\epsfig{file=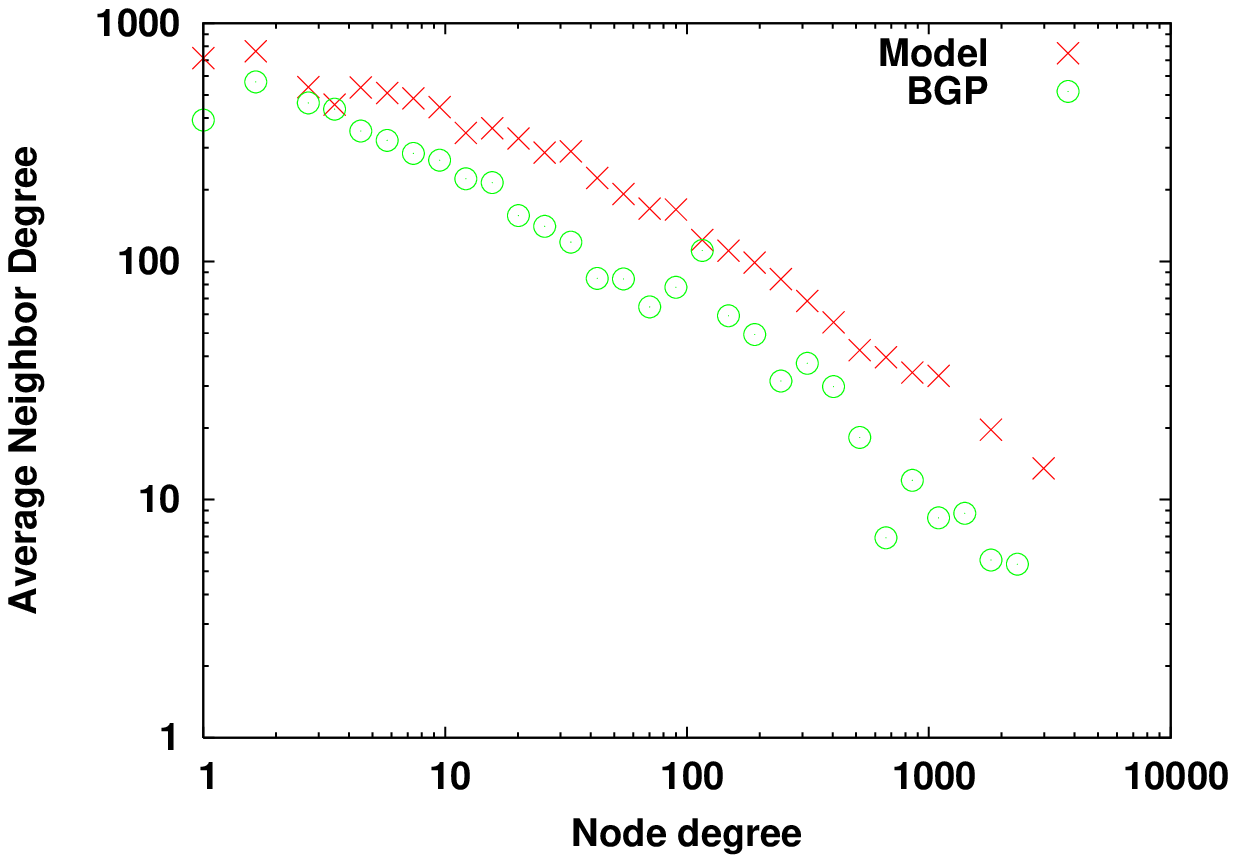,width=2.6in}
\label{fig_2_avg_nbr_c2p}
}
\subfigure[Node total degree vs.\ the average degree of its peers]{
\epsfig{file=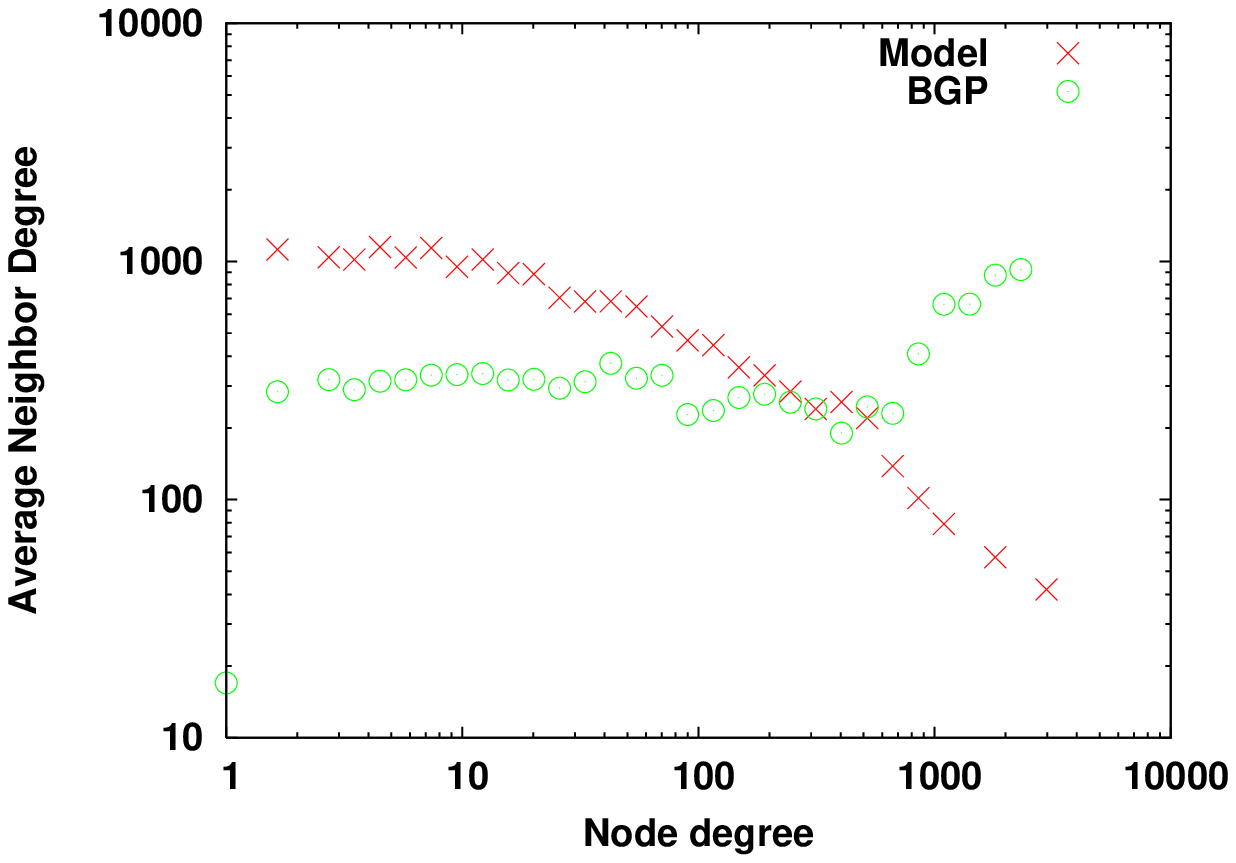,width=2.6in}
\label{fig_2_avg_nbr_peer}
}
\caption{Validation of the MPA model by simulation.}
\label{fig_correlations}
\end{figure*}

We have developed a model that describes the evolution of the
AS-level topology and validated the analytical results using
measured parameters. We now simulate the MPA model using all of the
measured parameters.

The MPA model generates \emph{annotated} graphs, with links connecting either
customers to providers (c2p links) or peers to peers (p2p links). Therefore
the total node degree is a sum of the degrees of three types---the
numbers of customers, providers, and peers attached to a node.
Dimitropoulos {\em et al.}~\cite{DiKr09} have
shown that the $2K$-annotated distribution
of the Internet essentially defines its structure. In other words,
if one randomizes the Internet
preserving its $2K$-annotated distribution, then the
randomized topologies will be almost identical to the original
Internet topology. The $2K$-annotated distribution is
a generalization of the joint degree distribution for graphs
with links annotated by their types. These types can be abstracted
by colors, and the traditional scalar node degree becomes a vector
of colors specifying how many links colored by what colors are attached
to the node. In the Internet case, these colors are {\em customer},
{\em provider}, and {\em peer}. The $2K$-annotated distribution is then
the joint distribution for the vectors of colored degrees of nodes
connected by differently colored links.

Given the findings in~\cite{DiKr09}, in order to show that our model reproduces
the Internet structure, it suffices to compare the $2K$-annotated
distributions in simulated networks and the Internet. Unfortunately,
the $2K$-annotated distribution is too multi-dimensional and sparse. Therefore,
we can work only with its projections. The reasonable and informative
projections include~\cite{DiKr09}:
(i)~the degree distribution~(DD): the traditional distribution
of total node degrees, i.e., the number of links of all types attached
to a node;  (ii)~the annotated distributions~(ADs): the
distributions of the number of customers, providers, and peers that
nodes have, i.e., the distribution of degrees of each type;
(iii) the annotated degree distribution~(ADD): the joint
distribution of customers, providers, and peers of nodes, measuring
the per-node correlations among these three degree types; and (iv)~the
joint degree distributions~(JDDs): the JDDs measure the correlations
of total node degrees ``across the links'' of different types.

\begin{figure}
\centering
\subfigure[Average neighbor degree]{
\epsfig{file=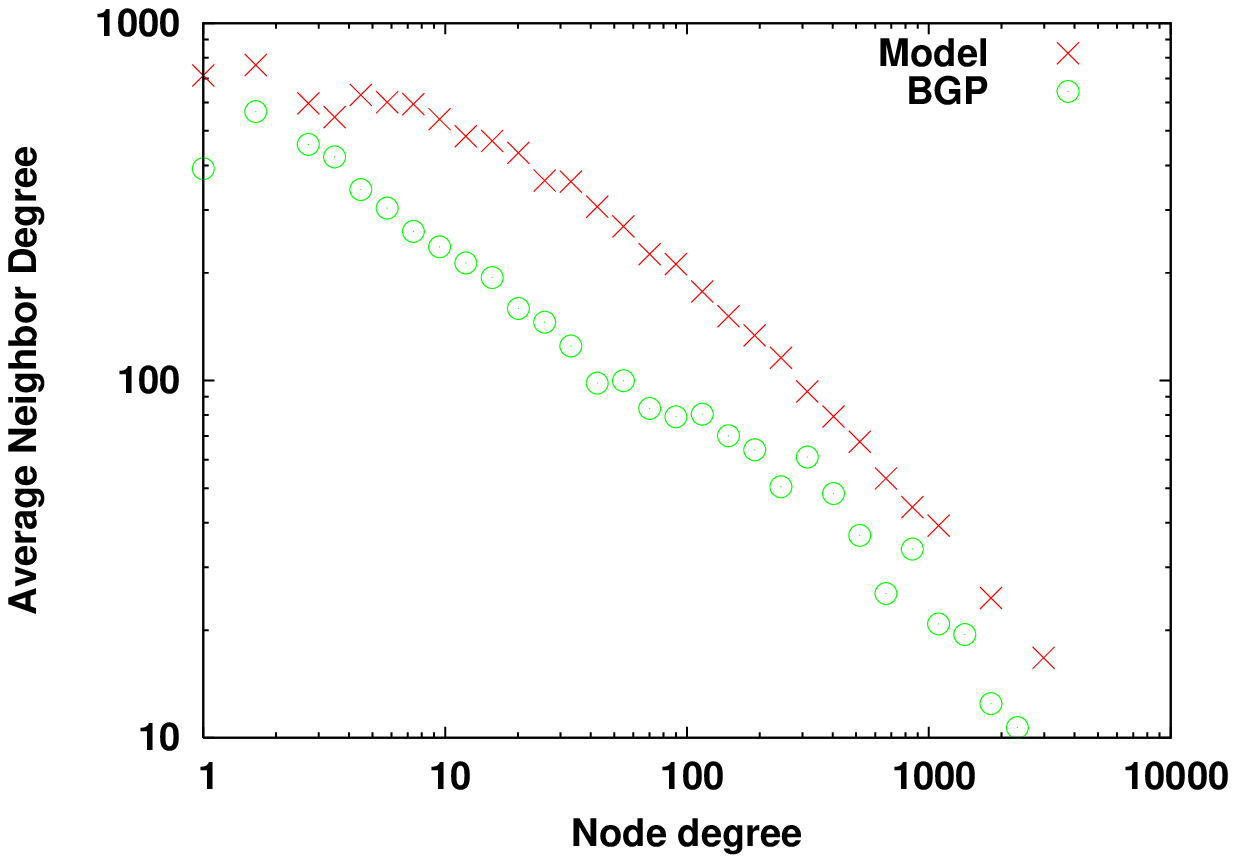,width=1.5in}
}
\subfigure[Clustering]{
\epsfig{file=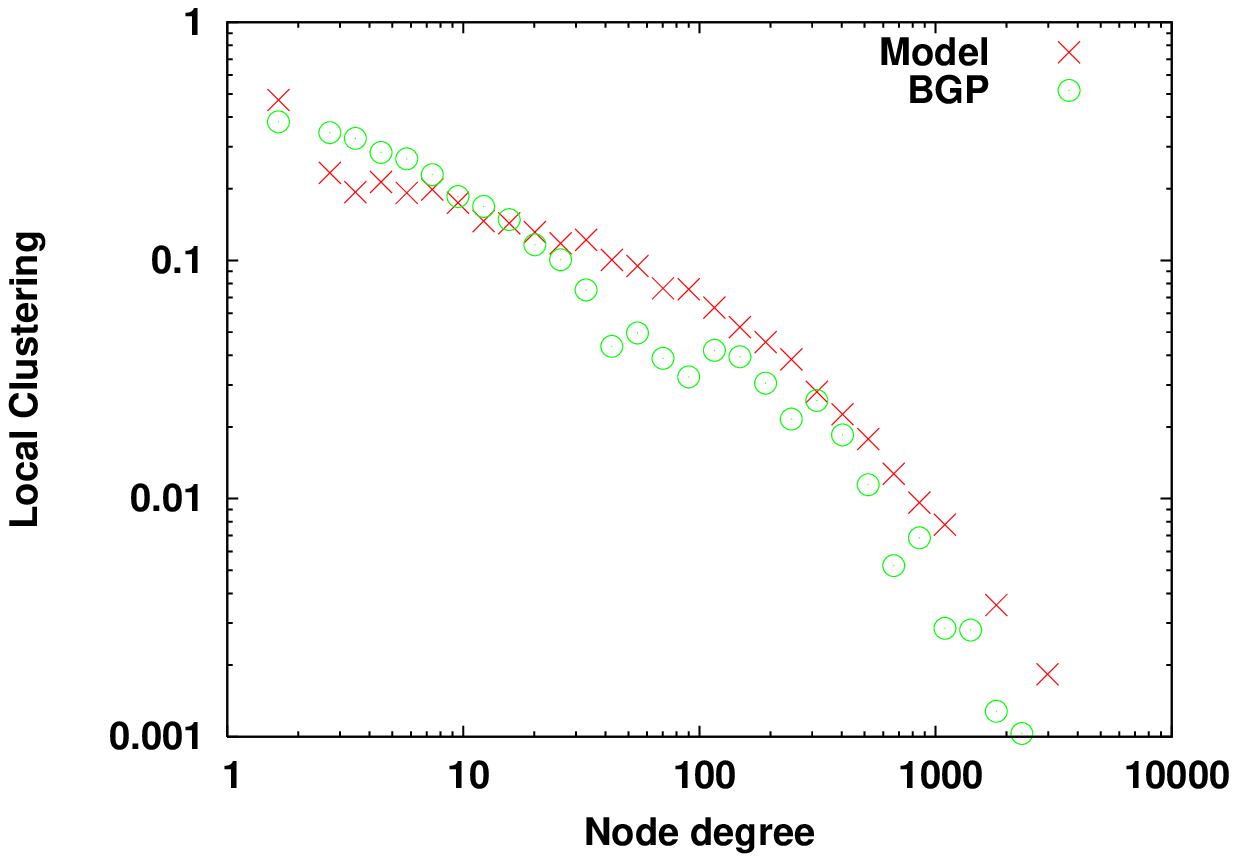,width=1.5in}
}
\caption{Average neighbor degree and clustering in simulations vs.\ real data.}
\label{fig_standard}
\end{figure}

Therefore, in our validation we compare all these metrics between the graphs that the MPA-model
produces and the Internet topology annotated with AS relationships
using~\cite{DiKrFo06} on Jan 1, 2007. The specific dataset used is
\url{http://as-rank.caida.org/data/2007/as-rel.20070101.a0.01000.txt},
which is a part of~\cite{as-rel-data} providing publicly available weekly
snapshots of the annotated Internet. These snapshots are based
on the Route Views BGP data~\cite{routeviews}.
The values of the parameters we use in our simulations are
$\rho=2.3$, $\nu=1$, $c=0.704$, and $m=1.86$, which we recall are
the ratio of the numbers of non-ISPs to ISPs, ratio of ISP
multihoming links to ISPs, ratio of peering links to ISPs, and the
average number of providers to which non-ISPs multihome,
respectively. We run the simulation with
deterministic link arrivals based on their arrival rates. The total
number of ISP and non-ISP nodes are $7,200$ and $16,800$.  These are
the numbers of nodes that we were able to classify in the dataset
using~\cite{DiKrRi06}. We do not model
bankruptcy since as mentioned earlier, the rate at which it occurs
is too small to get an accurate estimate of our bankruptcy ratio
$\mu$.

\textbf{Degree Distribution (DD):} Figure \ref{fig_all_ccdf} shows
the DD of the graphs generated by the MPA model, and its comparison
with the observed topology.  As predicted in the Section~\ref{sec_2nd_order}
the MPA model produces a power law DD, and the exponent of the CCDF
matches well the BGP data.

\textbf{Annotated Distributions (AD):} Figures
\ref{fig_cust_ccdf}--\ref{fig_prov_ccdf} show the ADs generated by
the MPA model.  We compare the customer, peer, and provider degree
distributions of the simulated graph with that of the BGP tables. As
predicted, the ADs of number of customers and peers are both power
law graphs with the same exponent as the DD.

We plot the CCDF of the number of providers that ISPs multihome to
(on linear $x$-axis and logarithmic $y$-axis) in
Figure~\ref{fig_prov_ccdf}.  They are approximately of form $1+X$,
where $X$ is exponentially distributed.  The curves show a
discrepancy in slope.  We believe that it arises due to the fact
that almost all the distribution mass is concentrated at small
degrees, as the mean is $2$, and the number of ISPs with high
multihoming degree is small.

\textbf{Annotated Degree Distribution (ADD):} Each ISP has some
numbers of providers, peers, and customers.  The ADD is the joint
distribution of these numbers across all observed ISPs. We illustrate
these correlations in Figures~\ref{fig_1_node_cust_vs_prov}
and~\ref{fig_1_node_peer_vs_prov}.  To construct these plots, we
first bin the ISP nodes by the number of providers that they have
(the $x$-axis), and then compute the average number of customers or
peers that the ISPs in each bin have (the $y$-axis). We observe that
the MPA model approximately matches the BGP data against these
metrics as well.

\textbf{Joint Degree Distributions (JDDs):} While the ADD contains
information about the correlations between the numbers of different
types of nodes connected to an ISP, it does not reveal information
about the degree correlations between the parameters of different
ISPs connected  to each other, i.e., whether higher degree ISPs are
more likely to peer with each other, etc. This information is
contained in the average neighbor connectivity, which is a summary
statistic of the joint degree distributions in Figures~\ref{fig_2_avg_nbr_c2p}
and~\ref{fig_2_avg_nbr_peer}. Specifically,
let the probability that a node of degree $k$ has a c2p link to a
node of degree $k'$ be called $P_{c2p}(k'|k)$.  Then the average
degree of the provider ISPs of ISPs that have degree $k$ is
$\bar{k}_{c2p}(k)= \sum_k' k' P_{c2p}(k'|k)$. In a full mesh graph
with $n$ nodes and undirected links, since all nodes have degree
$n-1$, the value of this coefficient is simply $n-1$.  We show the
normalized value  $k_{c2p}(k)/(n-1)$ in Figure
\ref{fig_2_avg_nbr_c2p}.  The similarly normalized values of
$\bar{k}_{p2p}(k)$ are shown in Figure~\ref{fig_2_avg_nbr_peer}.
These functions exhibit similar behaviors for the MPA model and BGP
data.

Coupled with observations in~\cite{DiKr09} that
the real Internet topology is accurately captured by its
$2K$-annotated distribution,
the results in this section provide evidence that the MPA model
reproduces closely the Internet AS-level topology across a wide range
of metrics. As an example, we show in Figure~\ref{fig_standard} two
standard topology metrics: the average neighbor degree and clustering
as functions of the total node degree. We observe that even though two-class preferential
attachment in Section~\ref{sec_2cls_prf_attach} produces tree networks,
its multiclass extensions in Section~\ref{sec_2nd_order}, implemented
in our simulations, closely reproduce clustering observed in the real
Internet, which is a consequence of the $2K$-annotated randomness of the Internet,
and the $2K$-annotated distribution match in Figure~\ref{fig_correlations}.
Finally, the average total node degree in the real Internet
and simulations is $4.1$ and $4.2$ respectively.

\section{Conclusion}\label{conclusion}

We constructed a realistic and analytically tractable model of the
Internet AS topology evolution that we call the multiclass
preferential attachment (MPA) model. The MPA model is based on
preferential attachment, and we believe it uses the {\em minimum
number of measurable parameters\/} altering standard preferential
attachment to produce annotated topologies that are remarkably
similar to the real AS Internet topology. Each model parameter
reflects a realistic aspect of AS dynamics. We measure all parameters using
publicly available AS topology data, substitute them in our derived
analytic expressions for the model, and find that it produces
topologies that match observed ones against a definitive set of
network topology characteristics. These characteristics are
projections of the second-order degree correlations, annotated with
AS business relationships. Matching them ensures that synthetic AS
topologies match the real one according to all other important
metrics~\cite{MaKrFaVa06-phys,DiKr09}.

The model parameter that has the most noticeable effect on the
properties of generated topologies reflects the ratio of ISP to
non-ISP ASs.  Contrary to common beliefs, the parameters taking
care of AS peering, bankruptcies, multihoming, etc., are
less important, as far the degree distribution is
concerned, although they do affect other network properties,
such as clustering.  No other parameters or complicated
mechanisms appear to be needed to explain the Internet topology
annotated with AS business relationships.  In other words,
preferential attachment, with the MPA modifications, appears to
explain the complexity of the AS-level Internet abstracted as an
annotated graph. An interesting open question concerns the origins
of preferential attachment in the Internet. Given that the vast
majority of AS links connect customer ASs to their
providers~\cite{DiKrFo06}, this question reduces to finding how
customers select their providers. The popularity of providers, their
``brand names,'' may be an important factor explaining the
preferential attachment mechanism acting in the Internet.

\begin{acknowledgements}

This work was supported in part by NSF CNS-0434996 and CNS-0722070, by DHS
N66001-08-C-2029, and by Cisco Systems.

\end{acknowledgements}


\end{document}